\documentclass[reprint,prl,aps,superscriptaddress,amsmath,longbibliography]{revtex4-1}
\usepackage{graphicx,txfonts,dcolumn}
\usepackage[colorlinks,linkcolor=blue,citecolor=blue,anchorcolor=blue,urlcolor=blue]{hyperref}

\begin{document}

\title{Explosive Percolation Obeys Standard Finite-Size Scaling in an Event-based Ensemble}

\author{Ming Li}
\email{lim@hfut.edu.cn}
\affiliation{School of Physics, Hefei University of Technology, Hefei, Anhui 230009, China}

\author{Junfeng Wang}
\affiliation{School of Physics, Hefei University of Technology, Hefei, Anhui 230009, China}

\author{Youjin Deng}
\email{yjdeng@ustc.edu.cn}
\affiliation{Hefei National Laboratory for Physical Sciences at Microscale and Department of Modern Physics, University of Science and Technology of China, Hefei, Anhui 230026, China}
\affiliation{MinJiang Collaborative Center for Theoretical Physics, College of Physics and Electronic Information Engineering, Minjiang University, Fuzhou, Fujian 350108, China}

\date{\today}

\begin{abstract}
Explosive percolation in the Achlioptas process, which has attracted
much research attention, is known to exhibit a rich variety of critical phenomena that
are anomalous from the perspective of continuous phase transitions.
Hereby, we show that, in an event-based ensemble, the critical behaviors in explosive percolation
are rather clean and obey the standard finite-size scaling theory,
except for the large fluctuation of pseudo-critical points.
In the fluctuation window, multiple fractal structures emerge
and the values can be derived from a crossover scaling theory.
Further, their mixing effects account well for the previously observed anomalous phenomena.
Making use of the clean scaling in the event-based ensemble, we determine with a high precision
the critical points and exponents for a number of bond-insertion rules,
and clarify ambiguities about their universalities.
Our findings hold true for any spatial dimensions.
\end{abstract}

\maketitle

Percolation is one of the paradigms in statistical physics
and probability theory~\cite{Stauffer1991}.
The standard percolation model on a lattice is defined
by randomly occupying sites or bonds with some
probability, and undergoes a continuous phase transition.
Simple alterations of the percolation, such as lattice type,
only result in different critical points,
and do not change the universality class~\cite{Stauffer1991}.
By adopting significantly different percolation rules,
such as rigidity percolation~\cite{Jacobs1995,Jacobs1996},
new universalities can arise.
Nevertheless, the continuity of the transition remains robust,
and the finite-size scaling (FSS) theory is always applicable.

In recent years, there has been an ongoing discussion on
the so-called Achlioptas process~\cite{Boccaletti2016,Saberi2015},
in which some intrinsic mechanism is introduced
to suppress the growth of large clusters. A basic way is called
the product rule~\cite{Achlioptas2009}.
At each time step, two empty bonds are randomly picked up,
the size-product of the two clusters containing the
ending sites of each bond is calculated,
and the one, leading to a smaller size-product, is inserted.
As a consequence, the onset of percolation is significantly
delayed, but once it happens, a large cluster emerges
suddenly, hence the name explosive percolation (EP).
EP has been observed in a wide
class of Achlioptas processes, including on regular lattices~\cite{Ziff2009,Ziff2010}
and scale-free networks~\cite{Cho2009,Radicchi2009}, and
in systems with other bond-insertion
rules~\cite{Friedman2009,Costa2010,DSouza2010,Nagler2011,Riordan2012}.
EP was perceived as a discontinuous transition when it was
introduced~\cite{Achlioptas2009,Friedman2009,Ziff2009,Cho2009,Radicchi2009,Ziff2010,Radicchi2010,DSouza2010,Cho2011},
but later studies suggested that the sharp transition
is continuous, despite displaying rich anomalous
behaviors~\cite{Costa2010,Lee2011,Grassberger2011,Riordan2011,Costa2014,DSouza2015}.

\begin{figure}[t]
\centering
\includegraphics[width=0.92\columnwidth]{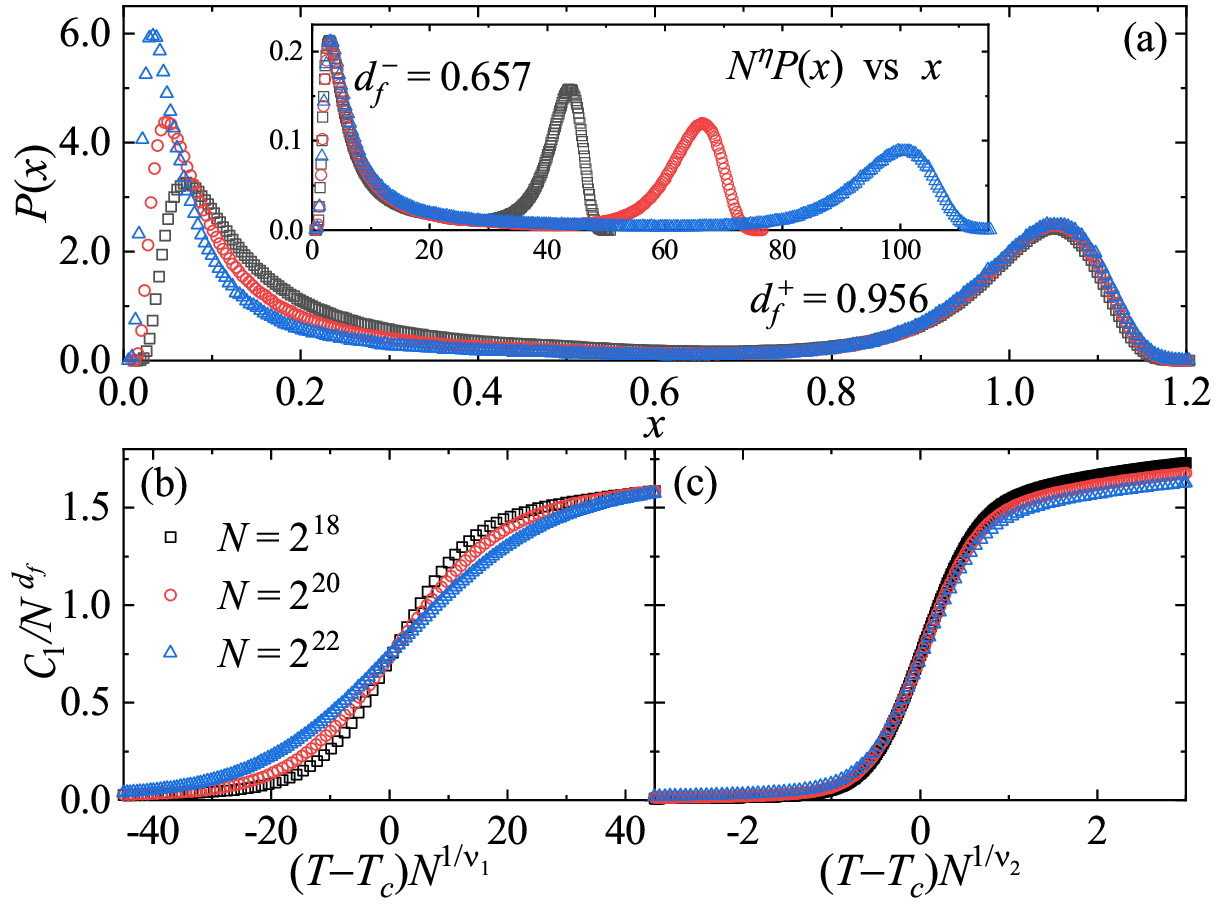}
\caption{Anomalous scaling behaviors in the conventional ensemble.
(a) The bimodal distribution $P(x)$ of the largest-cluster size $C_1$ at $T_c$.
Data collapse around the right peak is achieved by defining
$x=C_1/N^{d_f^+}$ with $d_f^+=0.956$,
while for the left peak, one has to use a smaller value $d_f^-=0.657$
and a rescaled exponent $\eta=0.08$.
(b, c) The scaling of $C_1$ near $T_c$,
with the exponents $1/\nu_1=0.740$ and $1/\nu_2=0.500$
(the correct fractal dimension $d_f=0.935$ is used here),
The data collapse is somewhat better for $\nu_2$,
which was incorrectly regarded as the correlation-length exponent~\cite{Grassberger2011,Fan2020}.
} \label{f1}
\end{figure}

Consider the largest cluster ${\cal C}_1$,
whose relative size, $m \! \equiv \! \langle {\cal C}_1 \rangle/N$
($N$ is the system volume), acts as an order parameter.
According to the FSS theory, at the critical point $T_c$,
$m$ scales as $N^{d_f-1}$, where $d_f$ is the fractal dimension
with respected to the system volume $N$~\footnote{
The standard definition of the fractal dimension $d_f^\ast$ is as $C_1\sim L^{d_f^\ast}$,
where $L$ is the length of hypercubic lattices.
Because of the absent of system length in random graphs,
we define the fractal dimension as $C_1\sim N^{d_f}$.
For a hypercubic lattice, we have $N=L^d$,
and thus $d_f=d_f^\ast/d$, where $d$ is the system dimension.}.
Further, the probability distribution of ${\cal C}_1$
can be renormalized to a single-variable function as
$P({\cal C}_1,N) \, d {\cal C}_1 = P (x) dx$,
with $x \equiv {\cal C}_1/N^{d_f}$.
However, as in Fig.~\ref{f1}(a) for random graphs,
EP displays a bimodal distribution~\cite{Grassberger2011,Tian2012},
and, further, multiple fractal dimensions
emerge--i.e., different values, $d_f^+$ and $d_f^-$,
are needed to collapse the data for different peaks.
Actually, neither of them is the correct fractal dimension,
as we shall show later.

The FSS theory also tells us that $C_1 \! = \! N^{d_f} \tilde{m}
(\delta T N^{1/\nu} )$, where $\delta T =T-T_c$,
$\nu$ is the correlation-length exponent with respected to the system volume $N$,
and $\tilde{m}(\cdot)$ is a universal function.
However, a wide range of $\nu$ values, inconsistent within the quoted errors,
has been reported for EP~\cite{Cho2010,Lee2011,Grassberger2011,Li2012,Fan2012}.
It was further observed~\cite{Grassberger2011,Fan2020} that there simultaneously
exists a pair of exponents, $\nu_1 \! < \! \nu_2 $, but neither of them
is sufficient to describe the scaling of $C_1$ data near $T_c$, see Fig.~\ref{f1}(b,c).
Other anomalous phenomena include
the powder-keg mechanism~\cite{Friedman2009},
non-self-averaging property~\cite{Riordan2012},
and hysteresis~\cite{Bastas2011}.
It seems that, despite being continuous,
EP does not obey the standard FSS theory,
and extracting correct exponents becomes difficult.
This leads to controversies about how the universality of EP
depends on bond-insertion rules.

By dynamically recording ${\cal C}_1(t)$,
where time step $t$ is also the number of inserted bonds,
the event, ${\cal T}_N \! \equiv \! t_{\rm max}/N$, can be located by the maximum point $t_{\rm max}$
of the incremental size, ${\cal C}_1(t)-{\cal C}_1(t-1)$~\cite{Manna2011,Lee2011,Nagler2011}.
Major progress was recently achieved~\cite{Fan2020}, in which
the pseudo-critical point $T_N \! \equiv \! \langle {\cal T}_N \rangle$
and the variance $\sigma_T^2 \! \equiv \! \langle \mathcal{T}_N^2\rangle-
\langle\mathcal{T}_N\rangle^2 $ are calculated.
The correct fractal dimension for random graphs, $d_f=0.935$, was obtained at ${\cal T}_N$.
It was further observed that the deviation decays
as $T_N-T_c\! \sim \! N^{-1/\nu_1}\!= \! N^{-0.75}$
but the fluctuation vanishes more slowly as $\sigma_T \! \sim \! N^{-1/\nu_2} \!= \! N^{-0.50}$.
The authors concluded that $\nu_2$ serves as the correlation-length exponent of EP.

In this Letter, we study EP in a similar way as in~\cite{Manna2011,Lee2011,Nagler2011,Fan2020,Feshanjerdi2021}.
A simple but important difference is that,
after locating ${\cal T}_N$,
the process was repeated according to the recorded sequence of inserted bonds.
This allows us to sample any quantity at any time step.
Here, we focus on two basic quantities--the order parameter $m$,
and the susceptibility $\chi \equiv \langle \sum_{i \neq 1} {\cal C}_i^2 \rangle/N$.
From the total number of clusters with size in $[s,s+\Delta s]$,
we also calculate the cluster-number density $n(s,N)$.
By definition, one has $\chi = \sum_s s^2 n(s,N)$.
We explore the scaling behaviors of these quantities, and
their dependence on the dynamic deviation
$\delta {\cal T} \equiv T-{\cal T}_N$.
To distinguish from the conventional ensemble of fixed bond density,
we call such dynamic sampling to be in the event-based ensemble.

We perform extensive simulations on random graphs and on hypercubic lattices in dimensions
from $2$ to $6$, and observe the following.
First, we find that, at ${\cal T}_N$ and
in terms of $\delta {\cal T}$,
the standard FSS theory applies well to any quantity as
$Q(T,N) = N^{Y} \tilde{Q}(\delta {\cal T} N^{1/\nu_1})$,
with $Y$ the associated exponent.
Note that the correlation-length exponent is unique,
which is $\nu_1$ instead of $\nu_2$.
Second, we reveal that the previously observed exponents $d_f^\pm$,
correspond to the fractal dimensions in the
fluctuation window ${\cal O}(N^{-1/\nu_2})$ at the super-
and sub-critical sides of ${\cal T}_N$, respectively.
Moreover, we propose a crossover scaling theory and
derive the values of $d_f^\pm$.
All these findings hold true for any dimension
and a number of bond-insertion rules.
Finally, we determine with a high precision
the percolation threshold and the critical exponents
for a number of bond-insertion rules, and identify their universalities.
For clarity, herein we only present the numerical results
for the basic EP (with the product rule) on random graphs,
and will publish other results elsewhere~\cite{Li_unpublished}.

\begin{figure}[t]
\centering
\includegraphics[width=1.0\columnwidth]{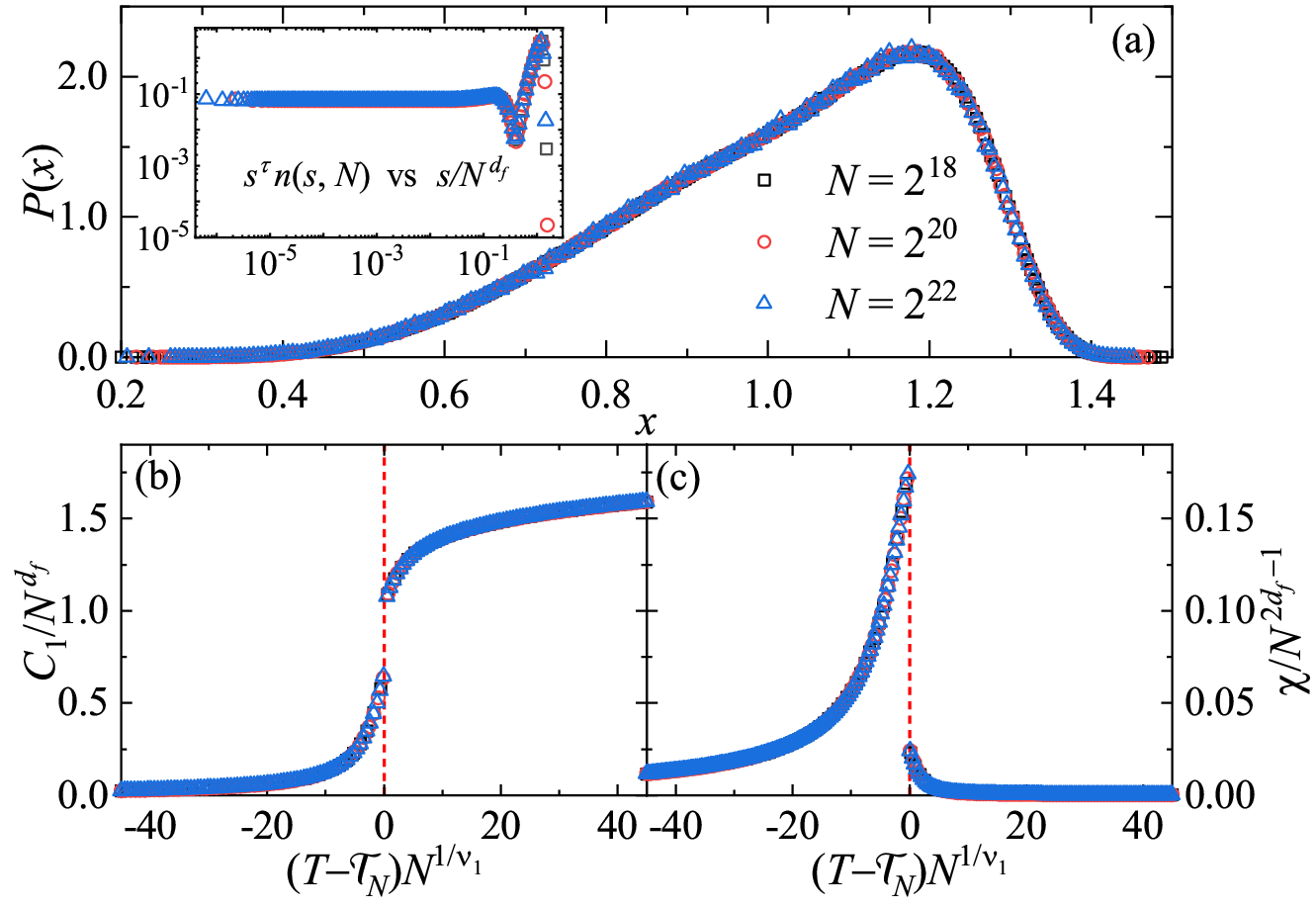}
\caption{Standard FSS behaviors in the event-based ensemble.
(a) At $\mathcal{T}_N$, $C_1$
has a uniform distribution $P(x=\mathcal{C}_1/N^{d_f})$,
where the correct fractal dimension is $d_f=0.935$ instead of $d_f^+$ or $d_f^-$.
The inset shows that the cluster-number density obeys
$n(s,N)=s^{-\tau}\tilde{n}(s/N^{d_f})$, with $\tau \! = \! 1+1/d_f\simeq 2.07$.
(b, c) Over a wide range at both sides of ${\cal T}_N$,
the standard FSS form holds well for $C_1$ and susceptibility $\chi$,
where the jump arises from the event-based definition of ${\cal T}_N$.
The correct correlation-length exponent is $1/\nu_1=0.740$,
instead of $1/\nu_2=0.500$~\cite{Grassberger2011,Fan2020}.
} \label{f2}
\end{figure}

\textit{Standard finite-size scaling in the event-based ensemble}.--
The probability distribution of $\mathcal{C}_1$ at ${\cal T}_N$,
is displayed in Fig.~\ref{f2}(a).
In contrast to Fig.~\ref{f1}(a), the distribution is smooth
and has a single peak, and, more importantly,
it can be expressed as a single-variable function as $P(x={\mathcal C}_1/N^{d_f})$.
Note that the correct fractal dimension, $d_f=0.935$,
equals neither to $d_f^+$ nor $d_f^-$.
In standard percolation, the cluster-number density at criticality
follows a power-law behavior up to a cutoff size $s_N \sim N^{d_f}$,
i.e., $n(s,N) =s^{-\tau}\tilde{n}(s/s_N)$,
and the Fisher exponent $\tau$ satisfies the hyperscaling relation
$\tau \! = \! 1+1/d_f$.
For EP, this gives $\tau=2.07$ from $d_f=0.935$,
and the nice data collapse in the inset of Fig.~\ref{f2}(a)
clearly demonstrates that $n(s,N)$ for EP obeys the standard FSS form.

Following the standard FSS ansatz,
we plot, respectively in Fig.~\ref{f2}(b) and (c),
the largest cluster $C_1$ and the susceptibility $\chi$
versus the renormalized dynamic deviation $z\!\equiv\!\delta{\cal T}N^{1/\nu_1}$,
where $d_f=0.935$ and $1/\nu_1=0.740$.
Excellent data collapse is achieved over a wide range of $z$,
which strongly supports that, despite of being sharp,
EP is a continuous transition and obeys the standard FSS theory.

To determine the percolation threshold $T_c$
and the critical exponents, $d_f$ and $1/\nu_1$,
we fit data to the standard FSS ansatz
\begin{eqnarray}
T_N &=& T_c+N^{-1/\nu_1} (b_0+b_1 N^{-\omega_1}+b_2 N^{-\omega_2}) \; ,
\label{eq:fit_TN} \\
C_1 &=& N^{d_f} (a_0+a_1 N^{-\omega_1}+a_2 N^{-\omega_2}) \; , \hspace{8mm}
(T={\cal T}_N)
\label{eq:fit_C}
\end{eqnarray}
where the terms with $\omega_i \, (i=1,2)$ are for finite-size corrections.
We obtain $T_c= 0.888\,449\, 1(2)$, $d_f=0.935(1)$ and $1/\nu_1=0.740(2)$,
where systematic errors have been taken into account.

\begin{figure}[t]
\centering
\includegraphics[width=1.0\columnwidth]{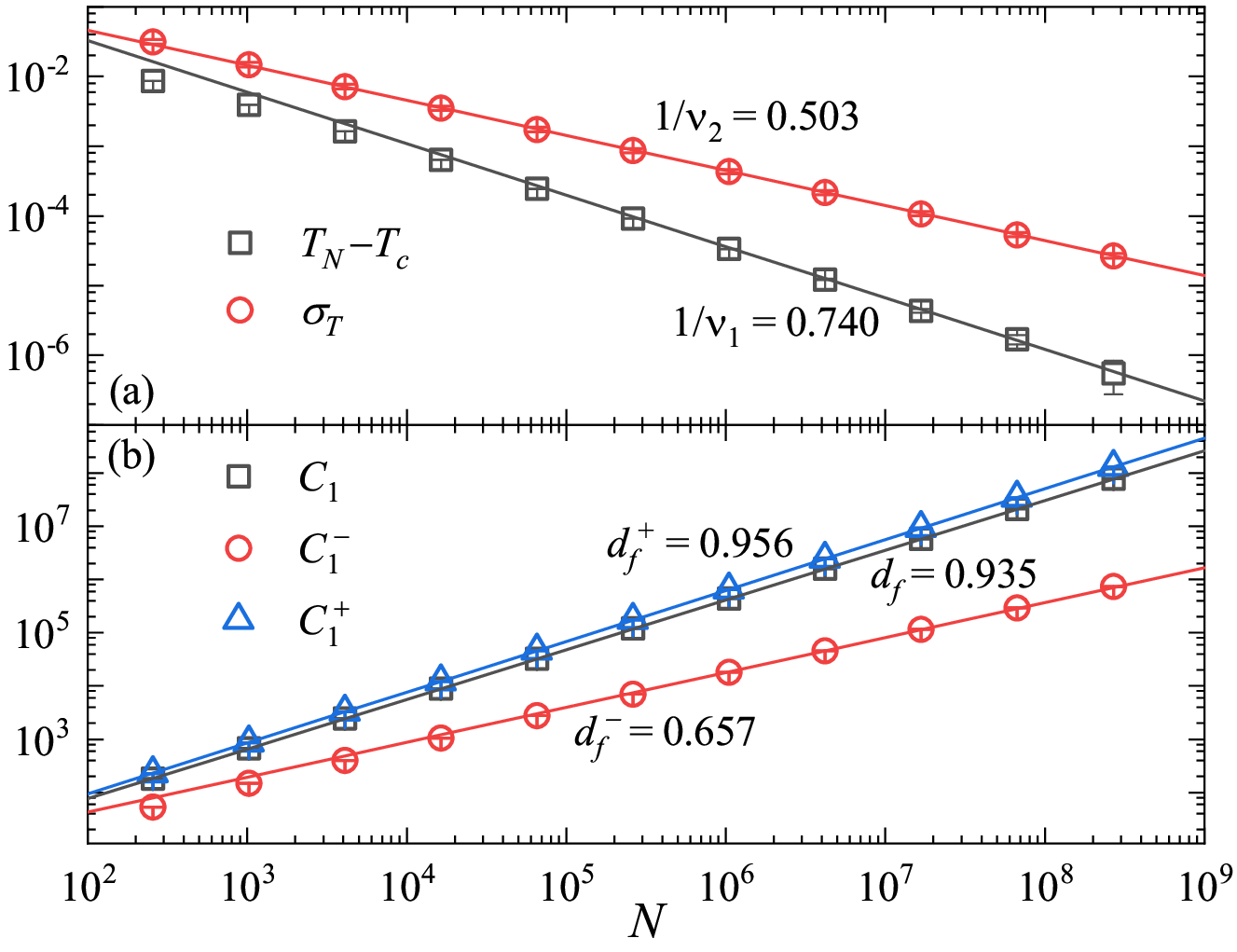}
\caption{Multiple critical exponents as determined by the standard FSS ansatz.
(a) The deviation is $T_N-T_c \! \sim \! N^{-1/\nu_1}$ with $1/\nu_1=0.740(2)$,
but the fluctuation is $\sigma_T \! \sim \! N^{-1/\nu_2}$ with $1/\nu_2=0.503(3)$.
The fluctuation window ${\cal O}(N^{-1/\nu_2})$
is larger than the standard scaling window $\mathcal{O}(N^{-1/\nu_1})$,
implying a non-self-averaging effect in the conventional ensemble.
(b) The largest clusters, $C_1$ at ${\cal T}_N$, $C_1^{\pm}$
in the fluctuation window ${\cal O}(N^{-1/\nu_2})$ for $T >{\cal T}_N$
and $T <{\cal T}_N$, respectively.
It is shown that, while $C_1$ has the fractal dimension $d_f=0.935(1)$,
$C_1^{\pm}$ have $d_f^+=0.956(2)$ and $d_f^-=0.657(3)$, respectively.
} \label{f3}
\end{figure}

\textit{Fluctuation window and multiple fractal dimensions}.--
For standard percolation, the deviation and the fluctuation of ${\cal T}_N$
are in the same order, $ T_N-T_c \! \sim \! \sigma_T \! \sim \! N^{-1/\nu_1}$,
where exponent $\nu_1$ is unique.
For EP, however, $\sigma_T$ vanishes
with a much slower speed and is governed by another exponent
as $\sigma_T \! \sim \! N^{-1/\nu_2}$~\cite{Fan2020}.
The fit of the $\sigma_T$ data gives $1/\nu_2=0.503 \approx 1/2$,
and the inequality,  $\nu_1 \!<\! \nu_2$, is clearly shown in Fig.~\ref{f3}(a).
Thus, beyond the standard scaling window ${\cal O}(N^{-1/\nu_1})$,
a fluctuation window ${\cal O}(N^{-1/\nu_2})$ is well defined.

We sample observables at $\mathcal{T}_N^\pm \equiv \mathcal{T}_N
\pm a N^{-1/\nu_2}$ and set $a=1$ for simplicity.
The largest-cluster sizes, $C_1^{\pm}$, are also well described
by a power-law scaling (Fig.~\ref{f3}(b)).
The fits by Eq.~(\ref{eq:fit_C}) give $d_f^+=0.956(3)$
for ${\cal T}_N^+$ and $d_f^-=0.657(3)$ for ${\cal T}_N^-$,
which agree well with those in Fig.~\ref{f1}(a).
This means that the two peaks in Fig.~\ref{f1}(a) actually
correspond to the scaling behaviors in the fluctuation window,
respectively at the super- and sub-critical sides.
It is thus revealed that the critical behaviors
in the conventional ensemble are effectively a mixture of those
in the fluctuation window.

\begin{figure}[t]
\centering
\includegraphics[width=1.0\columnwidth]{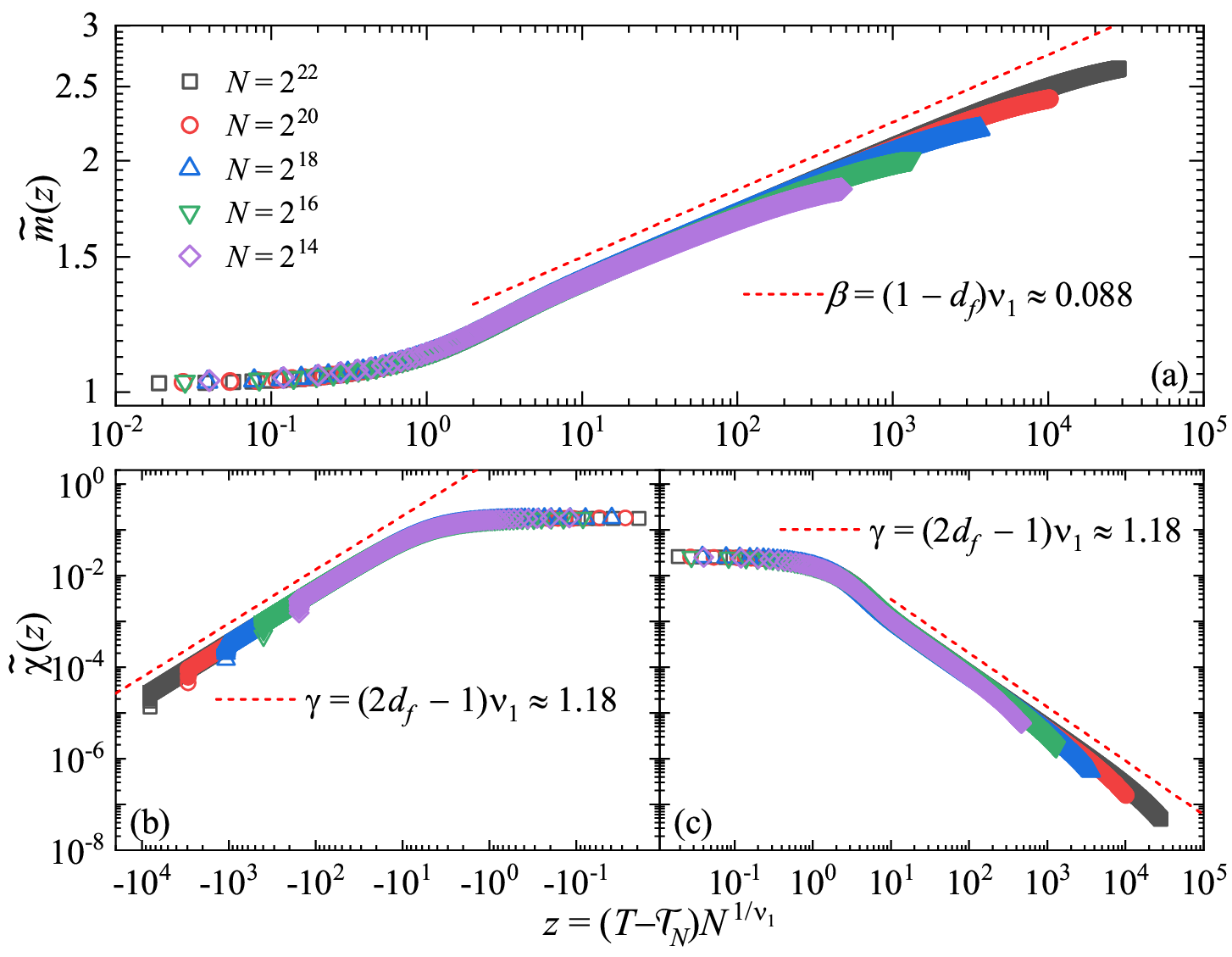}
\caption{Crossover scaling behaviors in terms of the renormalized dynamic
distance $z \! \equiv \! \delta {\cal T} N^{1/\nu_1}$.
(a) The universal function $\tilde{m}(z)$
in the FSS of the order parameter, $m \! = \! N^{d_f-1} \tilde{m}(z)$,
scales as $\tilde{m}(z \! \to \! \infty) \! \sim \! z\:^\beta$ in the super-critical side,
with $\beta \! = \! (1-d_f)\nu_1 \! \approx \! 0.088$.
(b, c) The universal function $\tilde{\chi}(z)$ in the FSS of the susceptibility,
$\chi \!=\! N^{2d_f-1} \tilde{\chi}(z)$, scales as $\tilde{\chi}(|z|\!\to\!\infty) \! \sim \!
|z|^{-\gamma}$ with $\gamma=(2d_f-1)\nu_1 \approx 1.18$,
which holds true at both the sub-critical ($z\!<\! 0)$
and the super-critical ($z\!>\! 0)$ sides of ${\cal T}_N$.
}
\label{f4}
\end{figure}

The pseudo-critical points ${\cal T}_N$ typically deviate away from
the thermodynamic point $T_c$
by an amount of $ {\cal O}(N^{-1/\nu_2})$,
while the correct critical behaviors are around ${\cal T}_N$ within
a narrow window $ {\cal O}(N^{-1/\nu_1})$.
In terms of $ z \! = \! \delta {\cal T} N^{1/\nu_1}$,
the fluctuation window is infinitely large as $|z| \sim N^{1/\nu_1-1/\nu_2} \to \infty$.
This suggests that the mixing effect is over an infinite range,
and cannot be averaged out by taking more samples.
It is thus no surprising that anomalous critical phenomena arise at $T_c$.

\textit{Relation between multiple fractal dimensions}.--
From the scaling behaviors in Fig.~\ref{f2},
we expect that $d_f \!=\!0.935$ is the only correct fractal dimension
and $d_f^{\pm}$ can be derived.
According to the FSS theory,
the correlated size behaves as $\xi_s \! \sim \! |\delta {\cal T}|^{-\nu_1}$,
so that $|z| =|\delta {\cal T}| N^{1/\nu_1} \! \sim \! (N/\xi_s)^{1/\nu_1}$.
In the fluctuation window, which has $N \! \gg \! \xi_s$ from $z \!\to\! \infty$,
the thermodynamic scaling should be recovered.

Let us consider the crossover scaling from finite- to infinite-$N$.
For susceptibility, as $|z|$ increases,
$\chi \!=\! N^{2d_f-1} \tilde{\chi}(z)$
should gradually evolve to $\chi \! \sim \! |\delta {\cal T}|^{-\gamma}$.
To eliminate finite-$N$ dependence, it is requested that
$\tilde{\chi}(|z| \! \to \! \infty) \! \sim \! |z|^{-\gamma}$
with $\gamma/\nu \! = \! 2d_f-1$.
The thermodynamic correspondence of $C_1$ is
the order parameter $m \! = \! N^{d_f-1} \tilde{m} (z)$.
For a continuous phase transition and for infinite-$N$,
$m$ remains zero for $\delta {\cal T} \! < \!  0$ and
the long-range order is continuously developed as $(\delta {\cal T})\:^\beta$
for $\delta {\cal T} \! > \!  0$.
In the super-critical phase, the crossover scaling of $\tilde{m}(z)$
can be extracted as $\tilde{m}(z \! \to \! \infty) \! \sim \! z\:^\beta$
with $\beta/\nu \! = \! 1 \! -\!d_f$.
These are well supported by Fig.~\ref{f4}.

From $C_1 = N^{d_f} \tilde{m}(z)$, $\tilde{m}(z) \! \sim \! z\:^\beta$
and $ z \!\sim\! N^{1/\nu_1-1/\nu_2}$,
the $d_f^+$ value is readily calculated as $d_f^+=1-(1-d_f) (\nu_1/\nu_2) \approx 0.956$,
in excellent agreement with those in Figs.~\ref{f1}(a) and~\ref{f3}(b).
At the sub-critical side, from the correlated size
$\xi_s \! \sim \! |\delta {\cal T}|^{-\nu_1} \! \sim \! N^{\nu_1/\nu_2}$,
we expect $C_1^- \! \sim \! \xi_s^{d_f}
\! \sim \! N^{d_f (\nu_1/\nu_2)}$,
giving $d_f^- \!=\! d_f (\nu_1/\nu_2) \! \approx \! 0.632$.
This is somewhat smaller than $d_f^- \!=\! 0.657$ in Figs.~\ref{f1}(a)
and~\ref{f3}(b), and it can be explained by an alternative way
based on $\chi$ and $n(s,N)$.

Consider a sub-critical window ${\cal O}(N^{-1/\lambda})$ centered around $\mathcal{T}_N$,
with $\lambda \! > \! \nu_1$,
we have $\chi \!\sim\! N^{(2d_f-1) (\nu_1/\lambda)}$ from the crossover scaling of $\chi$,
and expect $n(s,N) \! = \! s^{-\tau}\tilde{n}(s/s_\lambda)$,
with the cutoff size $s_\lambda \! \sim \! N^{d_\lambda} \!<\! N^{d_f}$.
The number of clusters of size $s_\lambda$ is diverging,
which is $N_\lambda \! \sim \! N \: s_\lambda^{1-\tau} \!\sim\! N^{1- (d_\lambda/d_f)}$,
with $s^{1-\tau}$ for the cumulative cluster-number density.
With this, the leading term of $\chi$ can be expressed as
$s_\lambda^2 N_\lambda/N \sim N^{2d_\lambda-d_\lambda/d_f}$.
Thus, by setting $(2d_f-1)(\nu_1/\lambda) = 2d_\lambda-d_\lambda/d_f$,
the relation between $d_\lambda$ and $d_f$ is established $d_\lambda \!=\! d_f(\nu_1/\lambda)$,
and $d_f^-=d_f(\nu_1/\nu_2)$ is recovered for $\lambda \!=\! \nu_2$.
Note that $d_\lambda$ is to characterize
the typical size of a diverging number of clusters,
while $C_1$ is the largest one.
From the extreme-value theory, one expects $C_1 \!\sim\!
N^{d_\lambda} (\ln N)^\kappa$, where exponent $\kappa$ depends on
the distribution of cutoff clusters.
This explains why the fitting result ($d_f^-=0.657$)
is slightly larger than the predicted value ($d_f^-=0.632$).

\textit{Universalities}.--
Unlike in standard percolation, it is suggested that, for EP,
small alteration of bond-insertion rule can lead to
different critical exponents~\cite{Boccaletti2016}.
For instance, the basic product rule (PR) can be modified
into the sum rule (SUM)~\cite{Achlioptas2009}, which
calculates  the total size of the two clusters associated with each candidate bond.
Further, an additional rule (AD) can be adopted by preferentially
inserting the intra-cluster bond~\cite{Cho2011}.
One can also apply the best of $m$ rule, i.e.,
choose three candidate bonds ($m3$) or even more~\cite{Friedman2009}.
On random graphs, the rule of~\cite{Costa2010}, we call it
CDGM by combining the initials of the authors' surnames, is applied:
choose a pair of random sites and reserve the site in the smaller cluster,
repeat the procedure for the second pair,
and finally, insert a bond between the two reserved sites.
Controversies remain about how the EP universality depends on bond-insertion rules.
As an exemplified case, debate still exists whether the AD rule would change
the universality of EP~\cite{Cho2011};
the fractal dimension was even estimated to be larger than the system dimension,
which is clearly unphysical~\cite{Wu2022}.

\begin{table}[t]
\caption{Percolation thresholds $T_c$ and critical exponents for
various bond-insertion rules, including the product rule (PR),
the sum rule (SUM), the CDGM rule, the best of $m$ rule for $m=3$ (m3),
and the additional rule (AD). EP has two basic exponents,
the correlation-length exponent $\nu_1$ and the fractal dimension $d_f$,
and, in addition, it has the fluctuation exponent $\nu_2$.
It is argued that the fluctuation obeys the central-limit theorem and thus $\nu_2=2$ holds exactly.
} \label{t1}
\begin{ruledtabular}
\begin{tabular}{c|lll|l}
Rules   & \hspace{7mm}$T_c$  & \hspace{2mm}$1/\nu_1$
        & \hspace{2mm}$d_f$  & \hspace{2mm}$1/\nu_2$ \\
\hline
PR      & 0.888\,449\,1(2)   & 0.740(2)   & 0.935(1)  & 0.503(3)  \\
PR+AD   & 0.888\,449\,0(4)   & 0.740(3)   & 0.935(1)  & 0.504(3)  \\
SUM     & 0.860\,207(1)      & 0.80(3)    & 0.957(5)  & 0.503(2)  \\
SUM+AD  & 0.860\,206(1)      & 0.80(3)    & 0.953(5)  & 0.500(3) \\
CDGM    & 0.923\,207\,4(3)   & 0.8181(1)  & 0.9545(1) & 0.500(2) \\
$m3$    & 0.964\,789\,9(1)   & 0.875(1)   & 0.979(1)  & 0.501(1)
\end{tabular}
\end{ruledtabular}
\end{table}

In the event-based ensemble, we study EP for
a list of bond-insertion rules, and the results for random graphs
are given in Tab.~\ref{t1}. We obtain the following:
(1) The AD rule does not change the universality,
    or even the percolation threshold.
(2) Universalities are different for the PR, the SUM, and the $m3$ rule;
    the phase transition seems to be sharpest for the $m3$ rule.
(3) The CDGM rule seems to be in the same universality
    as the SUM rule, within the estimated errors. But its finite-size
    corrections are significantly smaller and the estimated exponents
    have much higher precision, which are in excellent agreement with
    the result of the numerical method~\cite{Costa2014}.

\textit{Discussions}.--
By an event-based method, we find that EP obeys the standard FSS theory.
As standard percolation, EP has two basic exponents, the fractal dimension $d_f$
and the correlation-length exponent $\nu_1$,
which can describe well the critical behaviors of any quantities
near the pseudo-critical points ${\cal T}_N$.
Nevertheless, EP has a large fluctuation of ${\cal T}_N$,
which is governed by another exponent $\nu_2 \! > \! \nu_1$.
This scenario holds true for different bond-insertion rules,
and for any dimension~\cite{Li_unpublished}.
The high-precision estimate of critical exponents enables us to
establish the EP universalities for various bond-insertion rules.

The obtained $\nu_2$ values agree well with $2$,
except for two dimensions where $1/\nu_2\!=\!0.484(4)$ is slightly smaller than $0.5$~\cite{Li_unpublished}.
In units of the renormalized dynamic deviation $z$,
the fluctuation of ${\cal T}_N$ is infinitely large $N^{1/\nu_1-1/\nu_2}$,
implying that the central-limit theorem is satisfied.
Thus, the fluctuation may asymptotically be of Gaussian type
and $\nu_2=2$ holds exactly.
On this basis, we argue that $\nu_2$ is merely a fluctuation exponent
and cannot act as a correlation-length exponent.

The anomalous phenomena in the conventional ensemble are revealed to be a mixture
of critical behaviors over the fluctuation window.
Since it is infinitely wide in units of the renormalized deviation,
the self-averaging effect is lacking, and
this leads to the inequivalence of different ensembles.
Moreover, the multiple fractal dimensions are derived
based on the crossover scaling from finite- to infinite-$N$.

The effective event-based method can find broad applications,
since large sample-to-sample fluctuations can widely exist in
systems like disordered ones~\cite{Bernardet2000}.
Moreover, the proposed crossover scaling theory may provide
important insights for connecting critical behaviors in different ensembles.

The authors acknowledge helpful discussions with Peter Grassberger, Jingfang Fan, and Sheng Fang.
The research was supported by the National Natural
Science Foundation of China under Grant No.~12275263, and the National Key
R\&D Program of China (Grant No.~2018YFA0306501).

\bibliography{ref}

\begin{thebibliography}{34}
\bibitem [{Li_()}]{Li_unpublished}%
  \BibitemOpen
  \href@noop {} {}\bibinfo {note} {Detailed simulations and data analyses for
  different bond-insertion rules and different spatial dimensions are in
  progress.}\BibitemShut {Stop}%
\end{thebibliography}

\end{document}